\documentclass{article}

\usepackage{arxiv}

\usepackage[utf8]{inputenc} 
\usepackage[T1]{fontenc}    
\usepackage{hyperref}       
\usepackage{url}            
\usepackage{booktabs}       
\usepackage{amsfonts}       
\usepackage{nicefrac}       
\usepackage{microtype}      
\usepackage{cleveref}       
\usepackage{lipsum}         
\usepackage{graphicx}
\usepackage[numbers]{natbib}
\usepackage{doi}
\usepackage{lscape} 

\title{
    Motional representation; the ability to predict odor characters using molecular vibrations  
}

\newif\ifuniqueAffiliation

\uniqueAffiliationtrue
\ifuniqueAffiliation 
\author{\href{https://orcid.org/0009-0000-4254-7803}{\includegraphics[scale=0.06]{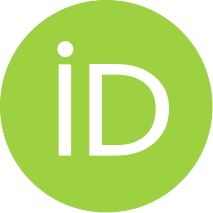}\hspace{1mm}Yuki Harada} \\
		Research and Education Institute for Semiconductors and Informatics Laboratory for Data Sciences of \\
		Kumamoto University, Kurokami 2-39-1, Chuo-ku, Kumamoto 860-8555, JP \\
	\texttt{yharada@kumamoto-u.ac.jp} \\
	\And
	\href{}{\hspace{1mm}Shuichi Maeda} \\
		Research and Education Institute for Semiconductors and Informatics Laboratory for Data Sciences of \\
		Kumamoto University, Kurokami 2-39-1, Chuo-ku, Kumamoto 860-8555, JP \\
	\texttt{Shuichi2503736@gmail.com} \\
	\And
	\href{https://orcid.org/0000-0003-4223-6735}{\includegraphics[scale=0.06]{fig/orcid.pdf}\hspace{1mm}Junwei Shen} \\
		Research and Education Institute for Semiconductors and Informatics Laboratory for Data Sciences of \\
		Kumamoto University, Kurokami 2-39-1, Chuo-ku, Kumamoto 860-8555, JP \\
	\texttt{jwshen@kumamoto-u.ac.jp} \\
	\And
	\href{https://orcid.org/}{\includegraphics[scale=0.06]{fig/orcid.pdf}\hspace{1mm}Taku Misonou} \\
		University of Yamanashi, Emeritus Professors, Takeda 4-4-37 \\
		Kofu, Yamanashi, 400-8510, JP \\
	\texttt{tmisonou@gmail.com} \\
	\And
	\href{https://orcid.org/}{\includegraphics[scale=0.06]{fig/orcid.pdf}\hspace{1mm}Hirokazu Hori} \\
		University of Yamanashi, Emeritus Professors, Takeda 4-4-37 \\
		Kofu, Yamanashi, 400-8510, JP \\
	\texttt{hirohori@yamanashi.ac.jp} \\
	\And
	\href{https://orcid.org/0000-0002-6437-6993}{\includegraphics[scale=0.06]{fig/orcid.pdf}\hspace{1mm}Shinichiro Nakamura} \\
		Research and Education Institute for Semiconductors and Informatics Laboratory for Data Sciences of \\
		Kumamoto University, Kurokami 2-39-1, Chuo-ku, Kumamoto 860-8555, JP \\
	\texttt{shindon@kumamoto-u.ac.jp} \\
}

\else
\usepackage{authblk}

\newbox{\orcid}\sbox{\orcid}{\includegraphics[scale=0.06]{fig/orcid.pdf}} 
\author[2]{%
	\href{https://orcid.org/0009-0000-4254-7803}{\usebox{\orcid}\hspace{1mm}Yuki Harada\thanks{\texttt{yharada@kumamoto-u.ac.jp}}}
}
\author[2]{%
	\href{https://orcid.org/}{\usebox{\orcid}\hspace{1mm}Shuichi Maeda\thanks{\texttt{Shuichi2503736@gmail.com}}}
}
\author[2]{%
	\href{https://orcid.org/0000-0003-4223-6735}{\usebox{\orcid}\hspace{1mm}Junwei Shen\thanks{\texttt{jwshen@kumamoto-u.ac.jp}}}
}
\author[3]{%
	\href{https://orcid.org/}{\usebox{\orcid}\hspace{1mm}Taku Misonou\thanks{\texttt{tmisonou@gmail.com}}}
}
\author[3]{%
	\href{https://orcid.org/}{\usebox{\orcid}\hspace{1mm}Hirokazu Hori\thanks{\texttt{hirohori@yamanashi.ac.jp}}}
}
\author[2]{%
	\href{https://orcid.org/0000-0002-6437-6993}{\usebox{\orcid}\hspace{1mm}Shinichiro Nakamura\thanks{\texttt{shindon@kumamoto-u.ac.jp}}}
}
\affil[1]{ \\
	Priority Organization for Innovation and Excellence Laboratory for Data Sciences, \\
	Kumamoto University, Kurokami 2-39-1, Chuo-ku, Kumamoto 860-8555, JP \\
}
\affil[2]{ \\
	Research and Education Institute for Semiconductors and Informatics Laboratory for Data Sciences of \\
	Kumamoto University, Kurokami 2-39-1, Chuo-ku, Kumamoto 860-8555, JP \\
}
\affil[3]{ \\
	University of Yamanashi, Emeritus Professors, Takeda 4-4-37 \\
	Kofu, Yamanashi, 400-8510, JP \\
}
\affil[4]{ \\
	Faculty of Advanced Science and Technology, \\
	Kumamoto University, Kurokami 2-39-1, Chuo-ku, Kumamoto 860-8555, JP \\
}
\fi

\renewcommand{\shorttitle}{
	A simple DNN regression for the chemical composition in essential oil 
}

\hypersetup{
pdftitle={
		Motional representation; the ability to predict odor characters using molecular vibrations  
},
pdfsubject={
	分子の動的な記述子の試み；匂いの質予測における分子振動の予測能の解析
},
pdfauthor={
	Yuki Harada, 
	Shuichi Maeda, 
	Junwei Shen, 
	Taku Misonou, 
	Hirokazu Hori, 
	Shinichiro Nakamura
},
pdfkeywords={odor, odorant, odor-sensing space, chemical space, molecular vibrations},
}

\begin{document}
\maketitle

\begin{abstract}

The prediction of odor characters is still impossible based on the odorant molecular structure. 
We designed a CNN-based regressor for computed parameters in molecular vibrations (CNN\_vib), 
in order to investigate the ability to predict odor characters of molecular vibrations. 
In this study, we explored following three approaches for the predictability; 
(i) CNN with molecular vibrational parameters, 
(ii) logistic regression based on vibrational spectra, and 
(iii) logistic regression with molecular fingerprint(FP). 
Our investigation demonstrates 
that both (i) and (ii) provide predictablity, 
and also that the vibrations as an explanatory variable (i and ii) and logistic regression with fingerprints (iii) show nearly identical tendencies. 
The predictabilities of (i) and (ii), depending on odor descriptors, are comparable to those of (iii). 
Our research shows that odor is predictable by odorant molecular vibration as well as their shapes alone. 
Our findings provide insight into the representation of molecular motional features beyond molecular structures. 

\end{abstract}


\section{Introduction}


\textit{Odorants chemical space} is the collection of volatilized chemical compounds 
that humans and many animals can perceive via their sense of smell. 
The total number of whole \textit{chemical space} is estimated to be $10^{60}$ 
that is the collection of all potentially possible exisiting compounds, including those yet to be found. 
\cite{reymond2015chemical, osolodkin2015progress} 
There could be almost 400,000 types of odor molecules in the odorant chemical space, in addition to the 20,000 known ones. 
\cite{mayhew2020drawing, mayhew2022transport, castro2022pyrfume} 
The prediction of perceived odor character / quality of odorants is an extremely challenging task based on their molecular structure. 
Because some odor character is not simply related directry to molecular structure, 
while others are related to molecular structure, 
such as aliphatic, aromatic, saturated, unsaturated, and polar.

There are different ways to represent chemical space, and there is no established implementation. 
\cite{capecchi2020one, wwwrdkit} 
Molecular fingerprint(FP) design is currently being refined, and there may be further approaches to detect molecular structure. 
\cite{boldini2024effectiveness} 
However machine learning with FP is a fundamental and/or irreplaceable approach 
for structure-activity relationship((Q)SAR) \cite{neves2018qsar,chastrette1997trends} and quantitative structure-odor relationship (QSOR). \cite{sanchez2019machine} 
In our previous study, 
we reported the relation between the odor descriptors and their structural diversity for odorants groups associated with each odor descriptor. 
Using the logistic regression with conventional FPs, we investigated 
the influence of structural diversity on the odor descriptor predictability. \cite{harada2024regression}

\begin{figure} 
    \centering
    \includegraphics[keepaspectratio, width=6.5 in ]{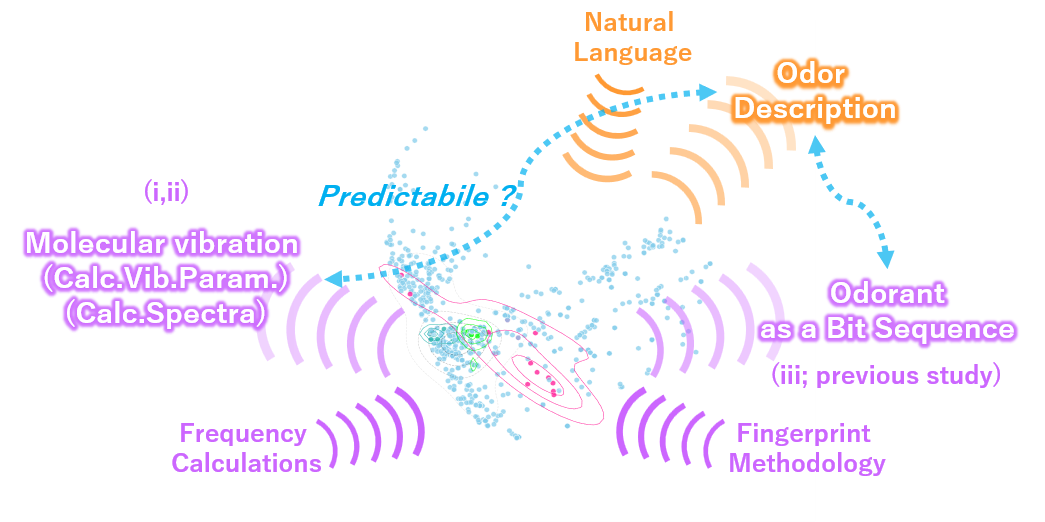}
    \caption{Comparison of three regressors to investigate the explanatory ability of molecular vibrations }
\end{figure} 

Since Richard Axel and Linda Buck discovered the role of G-protein-coupled receptors in olfactory cells in the 1990s, 
the biological mechanisms in the olfactory system have been successfully and thoroughly elucidated up to the present day. 
The primary stage of olfaction is the chemical interactions 
between olfactory receptors (around 400 types) and odorants (odor molecules, more than 20,000 types). \cite{mayhew2020drawing, mayhew2022transport, castro2022pyrfume}
The \textit{shape theory of olfaction}, 
meaning that odorant molecules are recognized by their shapes, is no doubt. 
The \textit{vibrational theory of olfaction}, 
which was a complementary previous hypothesis to it,
means that odorant molecular vibrations are responsible for olfaction rather than their shapes alone. 
While the \textit{shape theory of olfaction} is widely accepted concept of the olfaction, 
some researchers feel that both the form and vibrational properties of molecules have a role in olfaction. 
\cite{turin1996spectroscopic, block2015implausibility, vosshall2015laying}

In this study, we investigated 
whether the molecular vibration is effective 
and whether the current representation is sufficient for the molecular feature, 
as shown in Fig.1. 
We compared predictability of three regressors; 
(i) CNN\_vib with vibrational parameters, 
(ii) logistic regression based on vibrational spectra, and 
(iii) logistic regression with conventional FPs. 
For (i), we designed a CNN-based regressor with the quantum chemically computed parameters, we call it 'CNN\_vib' in this paper. 
For (iii), we already reported a study. \cite{harada2024regression}
As a natural extention of previous study, we evaluated differences in predictabilities of each regression from (i) and (ii). 
We will conclude and discuss the possibility of molecular vibration in odorants chemicals. 
Finally, we discuss focusing on the motional representation using molecular vibration beyond substructures.

\section{Materials and methods}

\subsection{Odorants and odor descriptors in the flavornet database } 

The data is collected from articles published since 1984 using GCO to detect odorants in natural products. 
It is available in pyrfume database 
\cite{castro2022pyrfume}
as well as Flavornet online database. 
\cite{acree2003flavornet}
It is a compilation of aroma compounds found in the human odor space.
The odorants are arranged by chromatographic and odor descriptors.

\subsection{Vibrational frequency calculation } 

Atoms in molecules vibrate relative to each other, including translations (external), rotations (internal), and vibrations. 
A diatomic molecule has only one motion, while polyatomic $N$ atoms molecules have $3N-6$ vibrations, known as normal modes. 
Each has a characteristic mode and frequency. 
Polyatomic molecules typically vibrate in the following modes: 
asymmetric, symmetric, wagging, twisting, scissoring, rocking and others.

All calculations, 
geometry optimizations and frequency calculations and the vibrational frequency calculation, 
were obtained by the Gaussian 16 suite of programs at the B3LYP/6-31G(d) level of theory. 
\cite{frisch2016gaussian}
We employed the following four parameters for all normal vibrational modes in this study : 
Harmonic frequencies ('F', cm**-1), 
reduced masses ('M', AMU), 
force constants ('C', mDyne/A) and 
IR intensities ('I', KM/Mole).

%

\subsection{Configuration of CNN\_vib } 

Fig.2 shows the regression with CNN\_vib
that we carried out in the current study; 
the explanatory variable is a matrix of four factors (F, M, C and I) for each vibrational mode. 
We call it 'Calc.Vib.Param.' in this paper. 
CNN\_vib is a multi-output regressor which gives multiple probabilities at once corresponding to each odor descriptor. 
The output is a matrix of two factors. 

\begin{landscape}
\begin{figure} 
    \centering
    \includegraphics[keepaspectratio, width=9.5 in ]{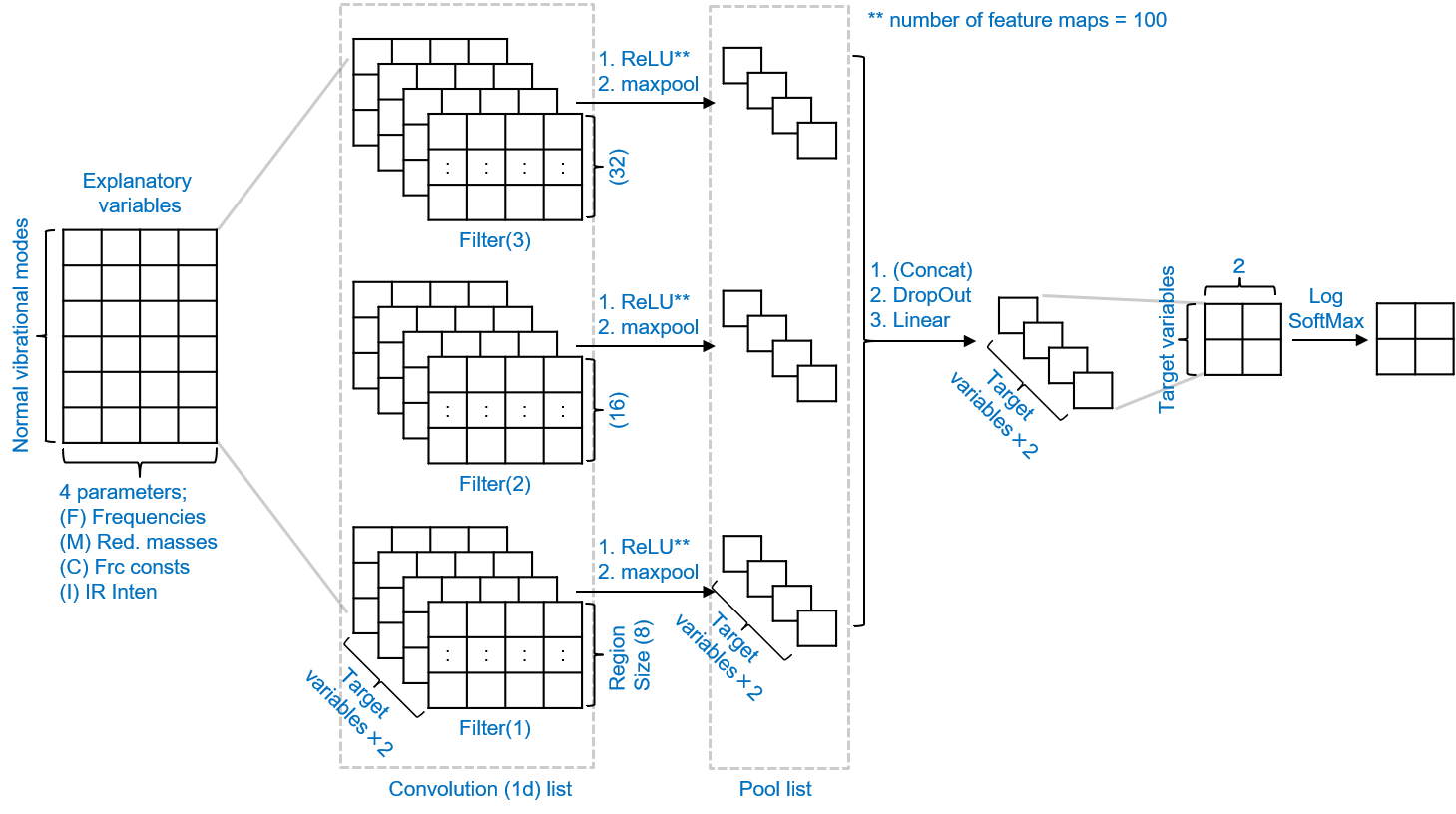}
    \caption{CNN\_vib design }
\end{figure} 
\end{landscape}

The predictability was evaluated using the Area Under the ROC Curve (AUC), 
which resulted in a value close to one, suggesting high classification accuracy.
We utilized \textit{k}-fold cross-validation to evaluate CNN\_vib. 
Then, the AUCs for each odor descriptor were calculated for training epoch numbers in CNN\_vib. 

We explored and optimized the design of CNN\_vib 
as well as their hyperparameters. 
The details of CNN\_vib design, 
such as the selection or sorting of explanatory variable matrix,  
hyperparameter tuning, 
the overlearning behavior and reproducibility are described in SI.


\subsection{Configuration of the logistic regression based on vibrational spectra } 

We also performed logistic regression based on vibrational spectra, which gives the probability of an odor descriptor.
The spectra were obtained 'F' and 'I' from 'Calc.Vib.Param.' after preprocessing. 
The preprocessing includes normalizing and 'moving sum', 
which is creating a series of sum of different selections of the full data set (see details in SI Fig.S003).
We call it 'Calc.Spectra' in this paper. 
The predictability was evaluated using the AUCs for each odor descriptor.
AUCs were obtained by iterative \textit{k}-fold cross-validation of the regression model for each odor descriptor, 
following the previous report.\cite{harada2024regression}

\subsection{Configuration of logistic regression with FPs } 

We performed logistic regression with four FPs, 
including MACCS keys, Extended-Conectivity Fingerprints (ECFPs), Avalon fingerprints, and RD-kit fingerprints (RDKFP), 
in the previous report.
\cite{harada2024regression} 
The predictability was evaluated using the AUCs. 
AUCs were obtained by iterative \textit{k}-fold cross-validation of the regression model, for each odor descriptor, for each FP. 




\section{Result and Discussion }


We investigated 
(i) CNN\_vib with 'Calc.Vib.Param.', 
(ii) the logistic regression with 'Calc.Spectra', and 
(iii) the logistic regression with FPs. 
The result concerning predictability in the three is summarized in Table 1. 
It is also visualized in Fig.3. 
The horizontal axis is the structural diversity, 
for which we adopted a conventional approach of Tanimoto similarity scores as 
an index in the current study. 
The vertical axis is the AUC for each odorants group.

\begin{table} 
    \centering
    \includegraphics[keepaspectratio, width=6.5 in ]{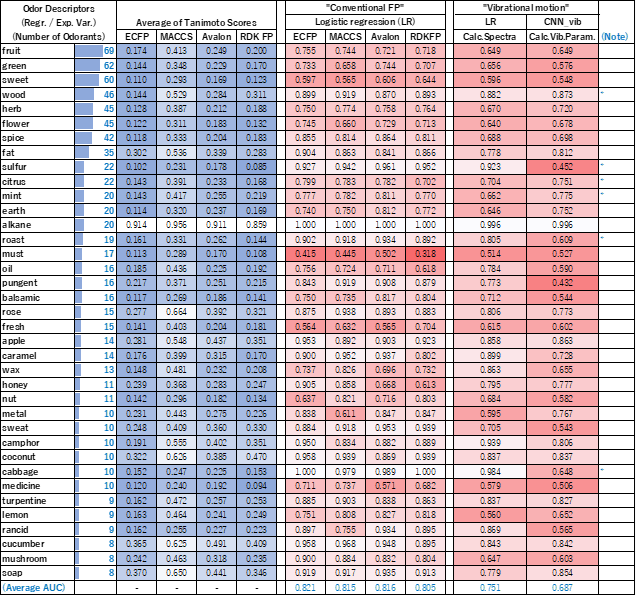}
    \caption{The AUCs by three regressor; 
    (i) CNN\_vib with calculated vibrational parameters ('Calc.Vib.Param.'), 
    (ii) the logistic regression based on vibrational spectra ('Calc.Spectra'), and 
    (iii) the logistic regression with four conventional FPs: MACCS keys, ECFPs, Avalon fingerprints, and RDKFP. 
    } 
\end{table} 

\begin{figure} 
    \centering
    \includegraphics[keepaspectratio, width=6.5 in ]{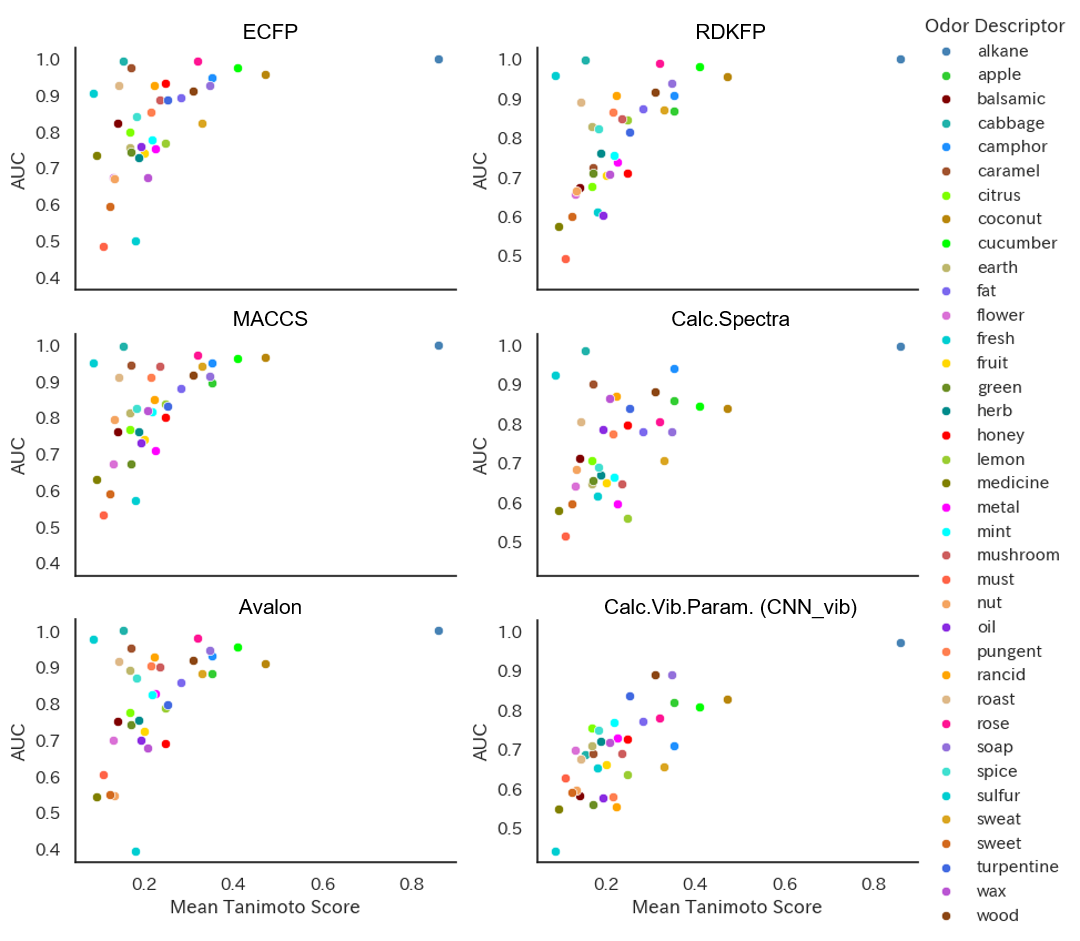}
    \caption{Similarity average vs AUC for each odorants group } 
\end{figure} 

In our previous study, 
we reported regression analysis and mapping based on the odorant chemical space, 
based on four conventional FPs: MACCS keys, ECFPs, Avalon fingerprints, and RDKFP. 
We found 
that the difficulties relates to the complexity included in the odor descriptor, 
and that the strong interplay is traced between structural diversity and the predictability of the odor descriptor, 
by all four FPs; the essential arguments are qualitatively conserved among the four fingerprints. \cite{harada2024regression}.

As a natural extention of previous study, in this study we evaluated differences in predictabilities of each regression from (i) and (ii). 
As a result, "wood" and "alkane", which had a high AUC in (iii), 
are also assigned with a high AUC in CNN\_vib.
In the same way, "must", "medicine" and "sweet", which had a small AUC in (iii), 
are also assigned with a small AUC in CNN\_vib. 
So far, we confirm the previously reported argument in three regressors. 
By contrast, we find a new feature obtained by CNN\_vib. 
It is noteworthy that the new feature is possible to be obtained only by CNN\_vib to be discussed in 3.1.


\subsection{Comparison of predictability in three regressors } 

At the simple CNN architecture, 
although we introduced a new algorithms different from logistic regression, 
the predictability of "wood", "citrus", and "mint" group by CNN\_vib is almost the same performance in comparison to that by logistic regression. 
By contrast, 
the predictability of "roast", "sulfur", and "cabbage" group by CNN\_vib is less than that by logistic regression as shown in Table 1 (see marked with *). 
It suggests that CNN\_vib may offer a qualitatively different perspective. 
We will discuss the differences in predictability at the  CNN\_vib depending on odorant group.

Fig.4 shows the examples of molecular structure in these six group. The following trends are observed. 
In the "wood" group (Fig.4(A)), hydrocarbon-sesquiterpenes are dominant having conventional functional groups such as ethers, alcohols, and ketones.
In the "citrus" group (Fig.4(B)), terpenes are dominant which have conventional functional groups such as alcohols and aldehydes.
The "mint" group (Fig.4(C)) contains menthol-like skeletons having conventional functional groups such as ketones, and also contains salicylic acid. 
Although the parent skeletons are diverse with various functional groups, even with the different functional groups, 
we infer that the similar vibrational modes can occur and give the high AUC by CNN\_vib. 
The "sulfur" group (Fig.4(D)) has distinctive S-containing functional groups. 
In the "roast" group (Fig.4(E)), pyrazines or thiazoles are dominant. 
The logistic regression with 'Calc.Spectra' for the "sulfur" and "roast" group 
show slightly inferior predictability to conventional FPs (see column 'Calc.Spectra' in Table 1).

This indicates that their structural feature may have a superior explanatory ability to the vibrational features for this case. 
The "cabbage" group (Fig.4(F)) has distinctive thioisocyanates or S-containing groups. 
Because the parent skeletons are diverse, with the characteristic functional groups on their skelton, 
the resulting 'Calc.Vib.Param.' vary widely. 
CNN\_vib shows only low predictability, because the vibrational modes are dissimilar (see also in Fig.S008-S013).
Their structure are diverse in the "sulfur" group and the "cabbage" group. 
Notice that high predictability is obtained through the regression with 'Calc.Spectra' processed with moving sum. 
Such a softening of vibrational characteristics is expected to produce good predictability. 

\begin{figure} 
    \centering
    \includegraphics[keepaspectratio, width=6.6 in ]{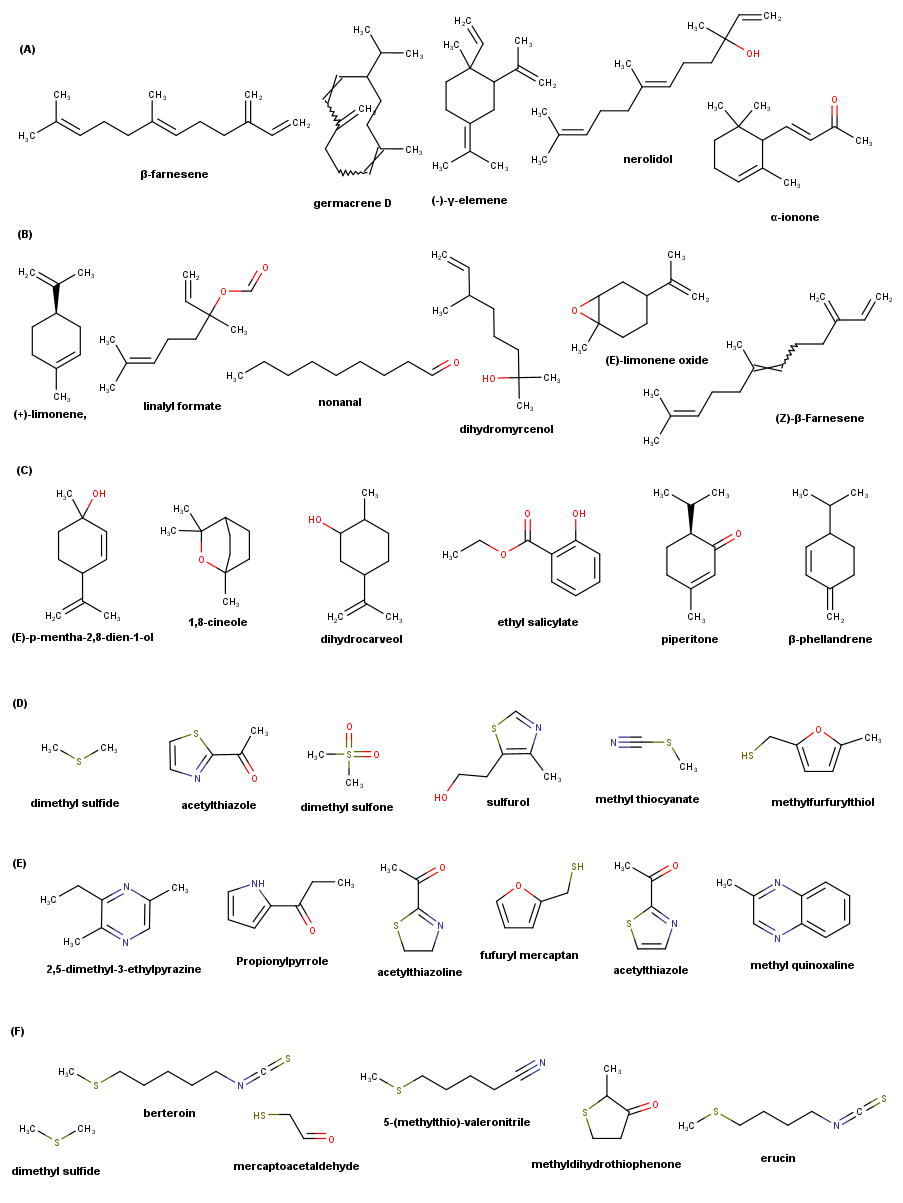}
    \caption{Example molecules in (A)"wood", (B)"citrus", (C)"mint", (D)"sulfur", (E)"roast", and (F)"cabbage" group} 
\end{figure} 

\subsection{Which descriptors demonstrated regression performance with molecular vibrations ?}

In the group "wood", "citrus" and "mint", 
because of their similar molecular structures as shown in Fig.4. 
it is potentially difficult that these molecules can be sniffed out in olfactory cognition. 
As a matter of fact, 
human will be able to distinguish their odor characteristics through repeated training. 
Then, it is a difficult question whether we can sniff out without training or not. 
It means that our culture has fine-resolutional cognition with using fine-resolutional \textit{evaluation word}, 
despite the strongly biased diversity for the three groups. 

On the other hands, the "roast", "sulfur" and "cabagge" groups have distinctive functional groups and can be smelled at a low threshold. 
Thus, they can be sniffed out in olfactory cognition and clearly apply these \textit{evaluation word}, leading to strong structure-odor relationship. 
Because of their structural diversity except their distinctive functional groups, 
CNN\_vib has only modest predictability for these groups. 
We had better consider the possibility that 
both cognitional and linguistic biased resolutional distributions must be taken into account 
in this regression study. 
This will be a very challenging subject in future for chemistry, data science and also for cognitive study.

%

\subsection{Insights into \textit{theory of olfaction} beyond molecular substructure } 

We attenpted to investigate whether the molecular vibration is truly irrelevant to their odor descriptors with two regressors; logistic regression and CNN\_vib. 
The \textit{shape theory of olfaction}, widely accepted concept of the olfaction, means that odorant molecules are recognized by their shapes, 
much like a key fits into a lock of olfactory receptor. 
Because these four FPs comprise of molecular substructures in a ligand molecule, 
they closely match the interaction and recognition of the \textit{shape theory of olfaction}. 
Our study demonstrated that 'Calc.Spectra' with logistic regression shows predictability, 
and that 'Calc.Vib.Param.' with CNN\_vib also shows predictability. 
We found that the molecular vibration has explanatory ability on odorant characters, rather than their shapes alone (not irresponsible for olfaction). 


\subsection{Insights into chemical space representation }

Some researches, inspired by the development of neural networks, in recent years have been published; 
image classification and segmentation, 
applications of Natural language Processing for line notations 
\cite{sharma2021smiles} 
and graph neural network (GNN) model.  
\cite{zhang2018deep, lee2022principal, qian2023metabolic, mortazavi2024recent, jacobs2025practical} 
Some regression studies indicated that 
the line notations model and graph neural networks performed even worse than image recognition-based models, \cite{sharma2021smiles} 
although their machine learning architectures are well-suited to chemical compounds. 
Even if their prediction performance has not yet meet expectations, those models have much rooms to be studied more profoundly.

CNN\_vib can read data in a variety of shapes and showed predictability: 
the explanatory variables, derived from vibrational frequency calculations, 
change shape according to their vibrational modes that correspond to the molecular complexity. 
It offers a fascinating alternative or complementary perspective, 
when the viewpoint shifts from static/spatial to motional or dynamic/temporal for the chemical space representation. 
In contrast, since the explanatory variable in conventional machine learning regression should have the same size of structure, 
reshaping of the parameters to spectrum is unavoidable for logistic regression. 
When investigating chemical space exploration with complex input or by nonlinear model, 
further deep learning techniques are to be studied. 
Our CNN\_vib demonstrated some explanatory ability of deep learning techniques on this problem. 
Such a techniques has the potential to improve the chemical space representation beyond molecular substructure and forecast physical attributes.

%
%

\section{Conclusions}

In this study, we investigated 
(i) the CNN with molecular vibrational parameters ('Calc.Vib.Param.'), 
(ii) the logistic regression based on vibrational spectra ('Calc.Spectra'), and 
(iii) the logistic regression with four conventional FP. 
Our investigation demonstrates 
that both (i) and (ii) provide predictablity and 
that (i), (ii) and (iii) show nearly the same tendencies in their predictability. 
It was also found that the predictabilities for some odor descriptors of (i) and (ii) are comparable to those of (iii). 
If the parent skeletons are diverse with the characteristic functional groups on their skeltons, 
their structural feature may have explanatory ability than the vibrational features, 
thus vibrational feature may not show predictability, 
because the resulting vibrational mode parameters vary widely. 
Our research demonstrated that the molecular vibration has explanatory ability on odorant characters. 
To improve the chemical space representation capable of predicting physical properties, 
our findings provide insight into the representation of molecular features beyond molecular substructure.


\section{ Data and Software Availability }

We used the RDkit \cite{wwwrdkit} 
for 
the FPs (MACCS, ECFPs, Avalon fingerprint and RDKFP)
. 
The regression methodogy, multi variate analysis and mapping is proprietary but not restricted to our program. 
The following Supporting Information is available; 
SourceAndSummary.zip (source and summary tables of regression results), 
out\_files.zip (Gaussian 16 output files for the odorants) and 
input\_files.zip (Gaussian 16 input files for the odorants). 


\section{List of abbreviations }

\begin{table} 
    \centering
    \caption{List of abbreviations }
    \begin{tabular}{|l|p{3.0 in}|} \hline
        Abbreviations & Description \\ \hline
        'F' & Harmonic Frequencies (cm**-1) \\ \hline
        'M' & Reduced Masses (AMU) \\ \hline
        'C' & Force Constants (mDyne/A)  \\ \hline
        'I' & IR Intensities (KM/Mole) \\ \hline
        ECFPs & Extended-Conectivity Fingerprints \\ \hline
        RDKFP & RD-kit Fingerprints \\ \hline
        IR & Infrared (Spectra) \\ \hline
        GCO & Gas Chromatography Olfactometry \\ \hline
        FP(s) & Fingerprint(s) \\ \hline
        AUC & Area Under the (ROC) Curve \\ \hline
        DNN & Deep Neural Network \\ \hline
        CNN & Convolutional Neural Network \\ \hline
        GNN & Graph Neural Network \\ \hline
        
    \end{tabular}
\end{table} 

\section{Acknowledgments}
The authors thank the anonymous reviewers for their valuable suggestions. 
This work was supported by Shorai Foundation for Science and Technology. 
The source code and related materials are available at our GitHub repository. \citep{ourrepo_github}


\bibliographystyle{unsrtnat}
\bibliography{reference}  






\end{document}